\documentclass[aps,prl,twocolumn,groupedaddress,showpacs]{revtex4}
\usepackage{graphicx}
\begin{document}

\title{Electron paramagnetic resonance detected via magnetization
measurements}


\author{K. Petukhov$^{1,2}$, W. Wernsdorfer$^1$, A.-L. Barra$^2$}

\affiliation{
$^1$Laboratoire Louis N\'eel, associ\'e \`a l'UJF, CNRS, BP 166, 38042 Grenoble Cedex 9, France\\
$^2$Grenoble High Magnetic Field Laboratory, CNRS, BP 166, 38042 Grenoble Cedex 9, France}


\date{February 5, 2005}

\begin{abstract}
Presented are magnetization measurements on a crystal
of Fe$_{8}$ single-molecule magnets using a Hall probe magnetometer.
Irradiation with microwaves at frequencies of 92
and 110-120 GHz leads to the observation of electron paramagnetic
resonance (EPR) detected via magnetization measurements.
A quantitative analysis of the results are introduced by means
of the spin temperature. It is shown that pulsed microwave 
experiments allow a better control over the spin excitation. 
\end{abstract}


\maketitle

Single-molecule magnets (SMM) are the final point in  the series
of smaller and smaller units from bulk matter to atoms. Up to now,
they have been the most promising candidates for observing quantum
phenomena because they have a well-defined structure with
well-characterized spin ground state and magnetic anisotropy.
These molecules can be regularly assembled in large crystals where
all molecules often have the same orientation. Hence, macroscopic
measurements can give direct access to single molecule properties.

High-frequency electron paramagnetic resonance (HF-EPR) has been
extensively employed to determine the magnetic anisotropy of 
SMMs~\cite{Barra96,Barra97,Hill98,Park02a,Hill02,Park02}. The
single-pass HF-EPR method~\cite{Barra96,Barra97} measures
resonance peaks corresponding to transitions between different
spin quantum levels. A complete set of resonance peaks at
different frequencies allows the determination of the spin
Hamiltonian parameters. A more sensitive cavity perturbation
HF-EPR technique~\cite{Hill98} allows in addition line shape
analysis~\cite{Park02a,Hill02,Park02}. The difficulties of these
HF-EPR spectroscopy techniques concern the control over the
electromagnetic environment of the sample. The use of overmoded
cylindrical resonators at high frequencies does not always provide
excited modes in compliance with an EPR geometry where the microwave
magnetic field is perpendicular to the applied magnetic field.
Undesirable instrumental effects like leaks, standing waves, and
amplitude-phase mixing~\cite{Mola00,Edwards04} can produce
unaccounted contributions to the HF-EPR spectrum and its
background, complicating consistent and straightforward line shape
analysis. Another issue is an inability to control the microwave
power exposed to the sample. Hence, the power dependence of line
widths and line shapes has not been qualitatively and systematically
studied.

Another powerful EPR tool for the SMM studies is the
frequency-domain magnetic resonance spectroscopy
(FDMRS)~\cite{Mukhin98, Mukhin01, VanSlageren03}.

In this communication we describe a complementary EPR method which
combines high-sensitivity magnetization measurements together with
microwave absorption measurements~\cite{Sorace03,WW_EPL04,Bal04,Barco_PRL04,Cage05}. 
The magnetization detection can be a
Hall-probe magnetometer~\cite{Sorace03,Barco_PRL04,Bal04},
a micro-superconducting  quantum interference devise
(micro-SQUID)~\cite{WW_EPL04}, a standard SQUID~\cite{Cage05} or
a vibrating sample magnetometer~\cite{Candela65}. The data of this
paper were obtained with a Fe8 single crystal placed into a
Hall-probe magnetometer with microwave radiation, thus having a
simple and affordable possibility for simultaneous magnetization
and EPR-like measurements. This approach 
obviates several experimental difficulties of cavity-employed
HF-EPR-spectroscopy mentioned above and controls better the
electromagnetic environment of the sample. Based on a Hall-bars
magnetometry, our technique exhibits an extraordinary sensitivity
being suitable for measurements of single micrometer-sized
crystals. Another advantage is that we can pulse microwave
radiation down to nanosecond time scales without dealing with a
lifetime of cavity modes at sub-millimeter wavelength frequencies,
where high sensitivity can be achieved only in expense of a high
Q-factor. When not limited by the microwave source and
waveguide cut-off frequencies, we are able to perform broadband
microwave measurements and employ concurrently two different
microwave frequencies for pulsed pump-and-probe measurements.
Finally, our technique allows us to get easily normalizable
spectra and introduce a quantitative analysis of the results by means
of the spin temperature.

The measurements were performed by using a magnetometer consisting
of several 10~$\times$~10~$\mu$m$^{2}$ Hall-bars~\cite{Sorace03}
on top of which a single crystal of Fe$_{8}$ was placed with an
easy axis approximately parallel to the magnetic field \textbf{B}.
The sample dimensions were 150 $\times$ 100 $\times$ 30
$\mu$m$^{3}$. The Fe$_{8}$ crystals were synthesized following
Weighart's method~\cite{Weighart84}. Note that much
smaller crystals could be used without loosing much sensitivity.
The magnetometer, placed into the commercial 16~T superconducting
solenoid, was combined with a microwave circuit consisting of a
continuous wave Gunn diode, an isolator, and a calibrated
attenuator. Pulsed radiation was achieved by
implementing a commercial SPST fast-PIN-diode switch with a
switching time of less than 3~ns. The microwave
radiation was guided and focused to the sample using an oversized
circular waveguide. Having two different Gunn-oscillators with
output power of 30~mW, we were able to perform measurements at
fixed frequency of 92~GHz and at several frequencies in the
frequency range of 110-120~GHz. The measurements were done in the
temperature range from 1.4 to 50~K, with a temperature stability of
0.05~K.

\begin{figure}
\includegraphics[width=3.5in]{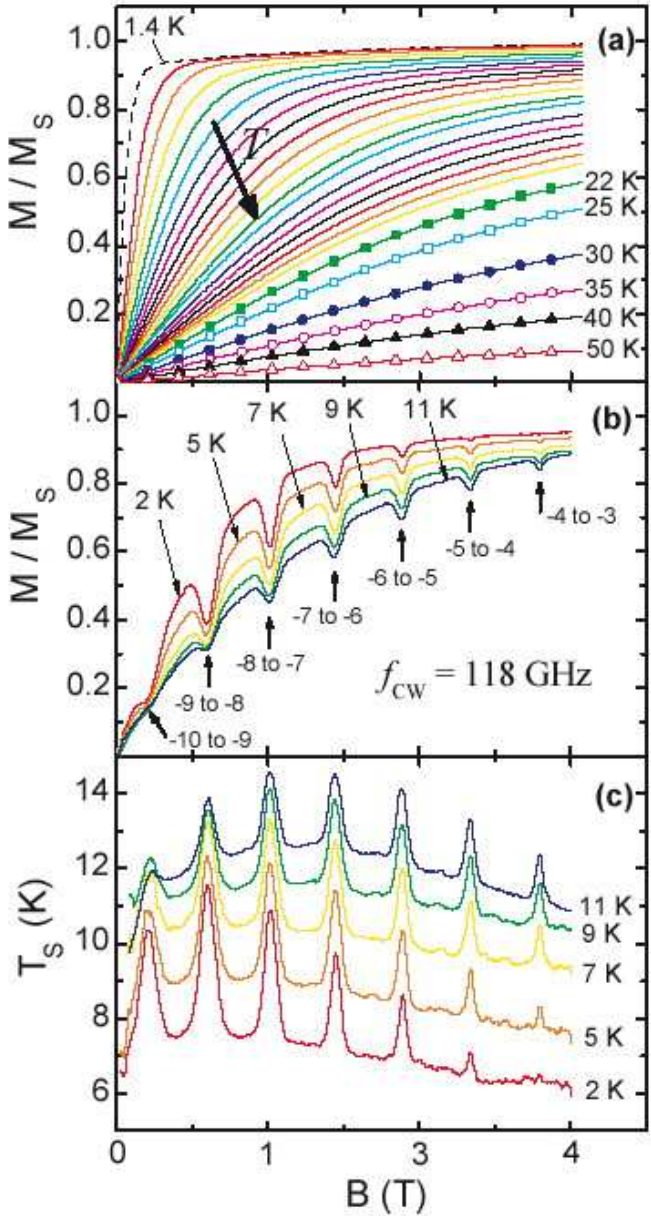}
\caption{\label{CW}(a) Magnetization of Fe$_{8}$ as a function of
magnetic field at different temperatures. 
The curves are normalized to the saturation 
magnetization value $M_{\rm s}$. The solid curves represent
the data measured from 2~K to 20~K in steps of 1~K. (b) Typical EPR-like
absorption spectra at different temperatures at continuous wave
frequency $f_{CW}=118$~GHz. (c) Spin temperature $T_{\rm S}$
\textit{versus} applied field $B$ at several cryostat temperatures
$T$, calculated using the mapping procedure described in the text.}
\end{figure}

Fig.~\ref{CW}a shows the temperature dependence of
magnetization of Fe$_{8}$ versus magnetic field 
at several temperatures from 1.4 to 50~K. 
When the sample is exposed to continuous microwaves (CW), the magnetization
curves show resonant absorption peaks, as depicted in
Fig.~\ref{CW}b for the frequency of 118~GHz. Similarly to
HF-EPR spectroscopy, absorption of microwave radiation takes place
at certain field values at a given frequency, when the microwave
frequency matches the energy difference between two energy states
with the quantum number $m_{S}$. The nearly evenly spaced
absorption peaks can be
attributed easily to the appropriate transitions (see
Fig.~\ref{CW}b).

\begin{figure}
\includegraphics[width=3.5in]{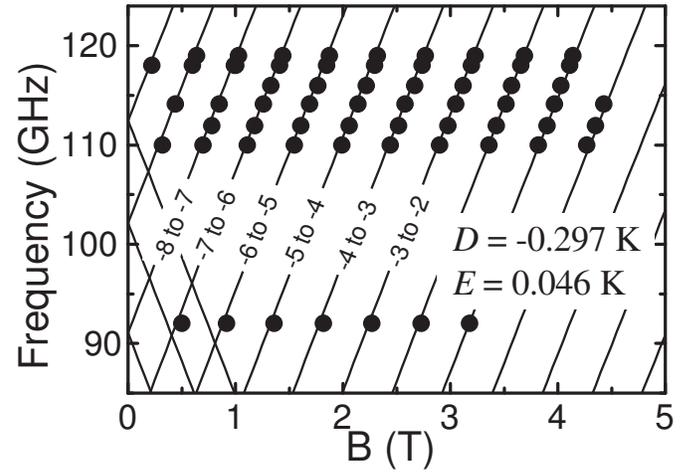}
\caption{\label{EPR} Field positions of the 
microwave absorption peaks at
several frequencies (dots). Solid lines 
represent the fit to the experimental data
obtained by diagonalizing the effective 
spin Hamiltonian~\cite{Abragam70,Barra96,Barra97}.}
\end{figure}

At 118~GHz the ground state resonance (transition from $m_{S}=-10$ to $-9$)
occurs close to zero field (at $B = 0.2$~T) and is
hardly visible on the slope of rapidly increasing magnetization. 
As magnetic field goes to zero, the magnetization also
goes to zero, and hence the sensitivity of detection of absorption
peaks goes to zero as well. Therefore, we need to perform a
transformation of the magnetization to a physical quantity which
does not depend on the magnetic field $B$.

Such a quantity can be obtained when the absorption
spectra~(Fig.~\ref{CW}b) are mapped on the magnetization
curves~(Fig.~\ref{CW}a) measured at different temperatures. 
For each magnetization point of the absorption spectra 
one finds, at the corresponding field, the temperature $T_{\rm S}$
that gives the same magnetization measured without 
microwave radiation~(Fig.~\ref{CW}a). The temperatures 
in between the measured once were obtained with an
interpolation. A typical result of such a mapping is
depicted in Fig.~\ref{CW}c. $T_{\rm S}$ can be called
{\it spin temperature} because the irradiation time is much
longer than the lifetimes of the energy levels of the
spin system which were found to be 
around 10$^{-7}$ s~\cite{WW_EPL00}.
The phonon relaxation time $T_{ph}$ from the crystal
to the heat bath (cryostat) is much longer
(typically between milliseconds and seconds~\cite{ChiorescuV15PRL00}).
The spin and phonon systems of the crystal are therefore
in equilibrium.

Fig.~\ref{EPR} shows the field positions of microwave absorption peaks
at several frequencies. These data allow the determination
of the crystal field parameters $D=-0.297$~K
and $E=0.046$~K of the effective spin
Hamiltonian of Fe8~\cite{Abragam70,Barra96,Barra97}.
Our result is very close to the values, obtained by
HF-EPR, inelastic neutron scattering (INS), and FDMRS
techniques~\cite{Barra96,Caciuffo98,Mukhin01}. 

\begin{figure}
\includegraphics[width=3.0in]{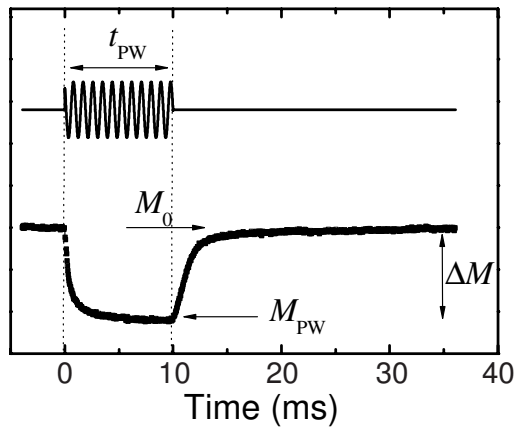}
\caption{\label{OSC}Typical oscillogram of a pulsed experiment. 
The magnetization was measured as a function of time
for a microwave pulse length of $t_{\rm PW}=10$~ms.}
\end{figure}

It is important to note that the obtained 
spin temperatures $T_{\rm S}$ are much
larger than the cryostat temperature $T$. This is
associated to a strong heating of the spin system, 
especially at low $T$. 
In order to reduce this heating
we need to perform low-power experiments.
The simplest way to reduce the power of CW microwaves
is to introduce an attenuator to the microwave
circuit. This solution reduces however the
sensitivity of absorption detection. A more advanced way 
is to use a pulsed microwave (PW) radiation.
In addition, this method might provide information
about the spin-lattice $T_{1}$ and spin-spin $T_{2}$
relaxation times.

\begin{figure}
\includegraphics[width=3.2in]{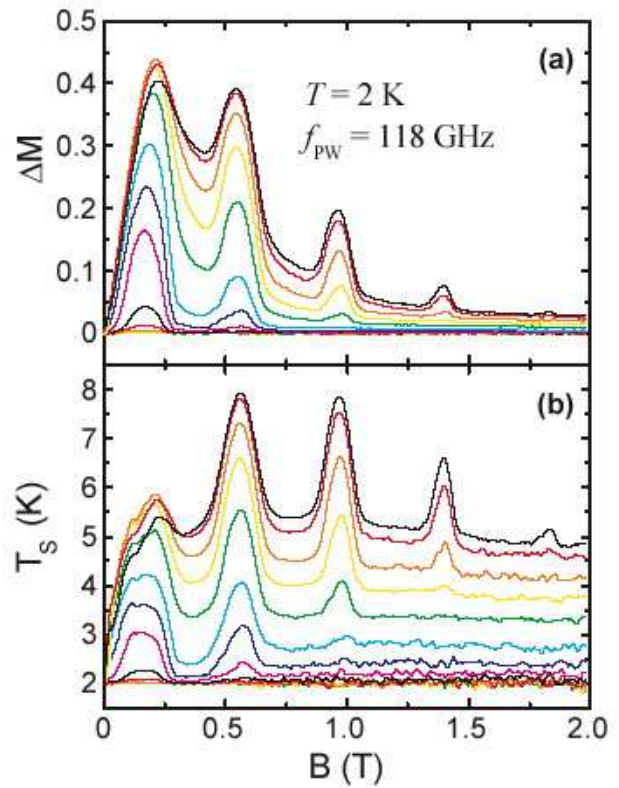}
\caption{\label{PW1}(Color online) (a) Magnetization variation
$\Delta M = M_{0} - M_{\rm PW}$ at the frequency of 118~GHz and 
the temperature of
2~K as a function of magnetic field. (b) Spin temperature as a function of
magnetic field calculated from $\Delta M$ in (a).
The pulse lengths in both figures are 10~ms, 5~ms, 2~ms, 1~ms, 500~$\mu$s,
200~$\mu$s, 100~$\mu$s, 50~$\mu$s, 20~$\mu$s, 10~$\mu$s, 5~$\mu$s,
2~$\mu$s, and 1~$\mu$s , from the
top to the bottom.}
\end{figure}

\begin{figure}
\includegraphics[width=3.3in]{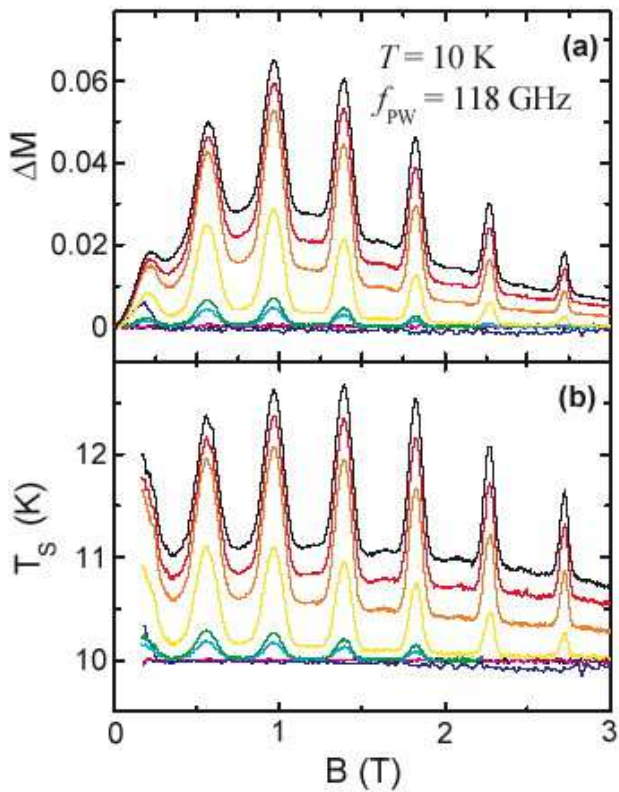}
\caption{\label{PW2}(Color online) (a) Magnetization variation 
$\Delta M = M_{0} - M_{\rm PW}$ at the frequency of 118~GHz and 
the temperature of
10~K as a function of magnetic field.  
(b) Spin temperature as a function of
magnetic field calculated from $\Delta M$ in (a).
The pulse lengths in both figures are 20~ms, 10~ms, 5~ms, 1~ms, 200~$\mu$s,
100~$\mu$s, 20~$\mu$s, 5~$\mu$s, and 1~$\mu$s from the
top to the bottom.}
\end{figure}

The microwave radiation was pulsed with a fast-PIN
switch with a switching time of less than 3~ns. The response time
of our magnetometer can also be optimized down to nanoseconds. 
The time evolution of the Hall voltage was detected with a fast digital
oscilloscope \textsc{Tektronix} TDS3054 with a 500~MHz bandwidth
and 5 GS/s sample rate. The scheme of the pulsed measurements
is depicted in Fig.~\ref{OSC}. 
The bottom part of Fig.~\ref{OSC} shows the data collected during
such an experiment with a pulse length of $t_{\rm PW}=10$~ms.
The magnetization before and 
at the end of the pulse has a value $M_{0}$ and $M_{\rm PW}$, respectively.
At the first milliseconds of the pulse,
the magnetization rapidly decreases and starts to saturate. 
A complete saturation is observed only for very long pulses
of several seconds.
After the pulse the magnetization increases back to the
initial value $M_{0}$. 
The time constants of activation $M_{0} \rightarrow M_{\rm PW}$
and consequent relaxation $M_{\rm PW} \rightarrow
M_{0}$ are connected to
the relaxation times $T_{1}$, $T_{2}$, and $T_{ph}$. Precise
time-resolved experiments are currently in progress.

Figs.~\ref{PW1}a and \ref{PW2}a show the difference
$\Delta M = M_{0} - M_{\rm PW}$ as a function of magnetic 
field at temperatures of 2 and 10~K. 
The pulse length was varied from 1~$\mu$s to 20~ms.
In contrast to the
CW experiments, the PW method can successfully resolve absorption
peaks near zero field. Analogous to CW experiments, the PW data
can be converted to the spin temperature $T_{\rm S}$ 
(Figs.~\ref{PW1}b and \ref{PW2}b).

The spin temperature curves show several interesting 
features. First of all,
the obtained spin temperatures $T_{\rm S}$ are much
closer to the cryostat temperature $T$ than for CW experiments. 
The peak positions of CW and PW are identical.
The line widths and shapes are depending on the 
pulse length.
At temperatures above 5~K
there are small but clearly pronounced peaks between the main absorption
peaks. Similar peaks were
observed in HF-EPR spectra of Fe$_{8}$ and 
ascribed to the presence of the $S=9$ excited state 
which is about 24 K above the $S=10$ ground state~\cite{Zipse03}.
The spin temperature curves show also a non-resonant background absorption
that was also seen by standard EPR methods~\cite{Park02a, Hill02, Park02}.
Our method might allow a quantitative investigation of this background.
One possible explanation of
this phenomenon implies low-lying spin states admixed
with the spin ground state, called S-mixing~\cite{Carretta04}. These
low-lying states can be thermally excited and thus they can
contribute to the observed background. 

In conclusion, we describe a complementary EPR method which
combines high-sensitivity magnetization measurements together with
microwave absorption measurements. This configuration allows
a quantitative analysis of the results by means
of the spin temperature.

This work was supported by the EC-TMR Network 
ÒQuEMolNaÓ (MRTN-CT-2003-504880), CNRS and Rhone-Alpe funding.
A. Cornia (University of Modena) is acknowledged for providing us with
Fe8 crystals, V. Mosser (Schlumberger, Utilities Technology Group) and
M. Konczykowski (Ecole Polytechnique) 
for fabricating the Hall probes. M. Dressel 
(University of Stuttgart) kindly loaned some of the equipment 
used for this study.


\end{document}